\documentstyle[psfig,referee]{mn2e} 
 
\def\etal{{\it et al. }} 
\title[Globular Clusters in NGC~3379] 
{Gemini/GMOS Spectra of Globular Clusters in the Leo Group Elliptical NGC~3379} 

\author[Pierce \etal] {
  Michael Pierce$^{1}$\thanks{mpierce@astro.swin.edu.au},
  Michael A. Beasley$^{2}$,
  Duncan A. Forbes$^{1}$,
   Terry Bridges$^{3}$,\\
\LARGE
  Karl Gebhardt$^{4}$,
  Favio Raul Faifer$^{5,6}$,
  Juan Carlos Forte$^{5}$,
  Stephen E. Zepf $^{7}$,\\
\LARGE
  Ray Sharples$^{8}$,
  David A. Hanes$^{3}$,
  Robert Proctor$^{1}$\\
  $^1$ Centre for Astrophysics \& Supercomputing, Swinburne University, 
  Hawthorn, VIC 3122, Australia\\ 
  $^2$ Lick Observatory, University of California, 
  Santa Cruz, CA 95064, USA\\ 
  $^3$ Department of Physics, Queen's University, 
  Kingston, ON K7L 3N6, Canada\\
  $^4$ Astronomy Department, University of Texas, 
  Austin, TX 78712, USA\\
  $^5$ Facultad de Cs. Astronomicas y Geofisicas, 
  UNLP, Paseo del Bosque 1900, La Plata, and CONICET, Argentina\\
  $^6$ IALP - CONICET, Argentina\\
  $^7$ Department of Physics and Astronomy, Michigan State University, 
  East Lansing, MI 48824, USA\\
  $^8$ Department of Physics, University of Durham, 
  South Road, Durham DH1 3LE\\
}

\begin{document} 
\maketitle 

\begin{abstract} 

The Leo Group elliptical NGC~3379 is one of the few normal elliptical
galaxies close enough to make possible observations of resolved
stellar populations, deep globular cluster (GC) photometry and high
signal-to-noise GC spectra.  We have obtained Gemini/GMOS spectra for
22 GCs associated with NGC~3379.  We derive ages, metallicities and
$\alpha$-element abundance ratios from simple stellar population
models using the multi-index $\chi^2$ minimisation method of Proctor
\& Sansom (2002).  All of these GCs are found to be consistent with
old ages, i.e. $\ga$10 Gyr, with a wide range of metallicities.  This
is comparable to the ages and metallicities Gregg \etal (2004) find
for resolved stellar populations in the outer regions of this
elliptical.  A trend of decreasing $\alpha$-element abundance ratio
with increasing metallicity is indicated.

The projected velocity dispersion of the GC system is consistent with
being constant with radius. Non-parametric, isotropic models require a
significant increase in the mass-to-light ratio at large radii. This
result is in contrast to that of Romanowsky \etal (2003) who find a
decrease in the velocity dispersion profile as determined from
planetary nebulae. Our constant dispersion requires a normal sized
dark halo, although without anisotropic models we cannot rigorously
determine the dark halo mass.

A two-sided $\chi^2$ test over all radii, gives a 2$\sigma$ difference
between the mass profile derived from our GCs compared to the
PN-derived mass model of Romanowsky \etal (2003). However, if we
restrict our analysis to radii beyond one effective radius and test if
the GC velocity dispersion is consistently higher, we determine a
$>3\sigma$ difference between the mass models, and hence favor the
conclusion that NGC~3379 does indeed have dark matter at large radii
in its halo.

\end{abstract} 
 

\section{Introduction}\label{sec_intro}

The Globular Cluster (GC) systems of numerous elliptical galaxies have
been well-studied photometrically (e.g. Gebhardt \& Kissler-Patig
1999; Kundu \& Whitmore 2001; Larsen \etal 2001). Since the discovery
of the bimodal colour distribution of GCs in elliptical galaxies (see
Harris 1999 for a review), several models have been proposed to
explain how these GC systems are formed.

Ashman \& Zepf (1992) propose that some, or all, of the
high-metallicity GCs will be formed during gas-rich merger events and
therefore be of a similar age to those mergers.  For the case of a
series of minor gaseous mergers, a roughly monotonic increase in
metallicity is expected with decreasing age.

The multi-phase collapse model of Forbes, Brodie \& Grillmair (1997)
proposes that the majority of GCs are native to their galaxy and
formed during two, or more, proto-galactic collapse phases.  Red,
metal-rich, GCs are expected to be 2--4 Gyrs younger than their blue,
metal-poor counterparts according to the semi-analytic GC formation
model of Beasley \etal (2002).

Cote, Marzke \& West (1998), invoke tidal capture of GCs from dwarf
galaxies to account for the blue GC population of large elliptical
galaxies.  A higher fraction of blue, metal-poor, GCs is expected
compared with spirals of a similar luminosity.  Assuming that blue GCs
form at the same time as their parent galaxy, ``down-sizing'' (Cowie
\etal 1996; Kodama \etal 2005) suggests that blue GCs captured from
dwarf galaxies should be younger than the native blue GCs.

Thus a key discriminant of the formation models is GC age.
Unfortunately from photometry alone it is very difficult to determine
the age and metallicity of individual GCs, due to the age-metallicity
degeneracy (Brodie \etal 2005), therefore high quality spectra are
required.  The stellar population properties of individual GCs can be
measured accurately with low-resolution integrated spectra of
sufficient signal-to-noise ($\geq$30).  The age-metallicity degeneracy
can be broken by measuring spectral Lick indices (Worthey 1994), and
comparing those indices to single stellar population (SSP) models.
High signal-to-noise spectroscopic studies (i.e. those capable of
determining ages for individual GCs) of elliptical galaxy GC systems
are limited to around a dozen galaxies (e.g. NGC~1399 Forbes \etal
2001; NGC~1023 Larsen \& Brodie 2002; NGC~1316 Goudfrooij \etal 2001;
NGC~3610 Strader \etal 2003, 2004; NGC~2434, NGC~3379, NGC~3585,
NGC~5846 and NGC~7192 Puzia \etal 2004; NGC~5128 Peng, Ford \& Freeman
2004; NGC~4365 Brodie \etal 2005; NGC~1052 Pierce \etal 2005).  Except
for Goudfrooij \etal and Peng \etal, these works find the majority of
GCs to be old ($>$10 Gyrs) with a small fraction of young GCs for some
galaxies.

In this work we report the analysis of Gemini/GMOS spectra of a sample
of GCs in NGC~3379. This is a typical, nearby (10.8$\pm$0.6 Mpc, Gregg
\etal 2004) E0/1 elliptical galaxy. It is of intermediate luminosity
(M$_V$=--21.06) with typical early-type colours, Mg$_2$ index and
velocity dispersion (Davies \etal 1987) for its luminosity. There is
no sign of any optical disturbance (Schweizer \& Seitzer 1992). There
is, however, a small nuclear dust ring at a radius of
1.5$^{\prime\prime}$ (van Dokkum \& Franx 1995) and some ionized gas
that extends to a radius of 8$^{\prime\prime}$ (Macchetto \etal 1996).

NGC~3379 is one of only a few elliptical galaxies to lie close enough
that its stars can be resolved by the Hubble Space Telescope (HST).
Resolved stellar population measurements using the {\it HST} NICMOS
camera in the {\it J} and {\it H} bands have been made by Gregg \etal
(2004).  Measuring individual stellar magnitudes and colours to just
below the Red Giant Branch (RGB) tip, they found the outer region
stellar population to be old, with ages $>$~8~Gyr and a mean
metallicity around solar. They noted similarities to the Galactic
bulge. For the central region, Terlevich \& Forbes (2002) used Lick
indices to estimate an age of 9.3 Gyrs, [Fe/H]=+0.16 and
[Mg/Fe]=+0.24.  This suggests that radial age and metallicity
variations are relatively small.

The {\it U, B, R, I} photometry of the GC system of this galaxy
reveals the classic GC colour bimodality (Whitlock, Forbes \& Beasley
2003).  Whitlock \etal estimate a GC specific frequency of S$_N$ =
1.1$\pm0.6$ which is low for an elliptical galaxy, although it is
important to remember that NGC~3379 belongs to a relatively small
group of galaxies (Leo) and therefore, a low S$_N$ is not unexpected
(Bridges 1992). The {\it B,V,R} photometry of Rhode \& Zepf (2004)
indicates a similarly low S$_N$ of 1.2$\pm$0.3 and that approximately
70\% of NGC~3379 GCs are blue.

To derive GC ages and metallicities we measure Lick indices from our
GMOS spectra. We then apply the multi-index $\chi^2$-minimisation
method of Proctor \& Sansom (2002), which employs all the available
Lick indices to break the age-metallicity degeneracy and
simultaneously measure the $\alpha$-element abundance ratios (see
Proctor, Forbes \& Beasley 2004 for an application to Galactic GCs and
Pierce \etal 2005 for NGC~1052 GCs).

In addition to the measurement of line indices, integrated spectra of
GCs allow the measurement of recession velocities, which confirm
whether the GCs are indeed associated with the galaxy being studied.
With a large enough sample of measured GC velocities it is possible to
probe the gravitational potential and therefore the dark matter halo
of a galaxy (e.g. Zepf \etal 2000; Cote \etal 2001; Cote \etal
2003; Peng \etal 2004; Richtler \etal 2004).

Measurements of planetary nebulae (PN) kinematics in NGC~3379 (Romanowsky
\etal 2003) reveal a decreasing velocity dispersion profile at large
radii. This suggested that NGC~3379 has a significantly lower
mass-to-light ratio (M/L = 7.1$\pm$0.6 at 5 R$_e$) than most large
ellipticals, indicating a minimal Dark Matter (DM) halo.  This result
would appear to be in conflict with the standard Cold Dark Matter
(CDM) galaxy formation picture in which all galaxies lie within
significant DM halos.  Here we use the measured velocities of our
sample of GCs, as well as the GC data of Puzia \etal (2004), as an
alternative data set to test this important claim.

In this paper we present our GMOS observations and data reduction of
GC spectra in Section \ref{sec_obs}.  An analysis of ages,
metallicities and $\alpha$-element abundance ratios derived from Lick
indices is presented in Section \ref{sec_age}.  In Section
\ref{sec_kin} we focus on GC kinematics and the implications for
the dark matter halo of NGC~3379.  We discuss our results in Section
\ref{sec_disc} and present our conclusions in Section \ref{sec_conc}.

\section{Observations and data reduction}\label{sec_obs} 

The observations described below are part of Gemini program
GN-2003A-Q-22.  GC candidates were selected from Gemini North
Multi-Object Spectrograph (GMOS; Hook \etal 2002) imaging, obtained
during January 2003, for three fields around NGC~3379. The data
reduction and GC candidate selection process is almost identical to
that of Forbes \etal (2004) for NGC~N4649, briefly, SExtractor (Bertin
\& Arnouts 1996) was used to select GC candidates based upon their
Gemini zero point colours (0.7$<g-i<$1.4), magnitudes (i $<$ 24) and
structural properties (i.e. objects with stellarity index $>$0.35).
Interstellar reddening towards NGC~3379 is E(B-V)=0.024 mag and is not
taken into consideration (Schlegel \etal 1998).  Spectroscopy of
NGC~3379 globular clusters were obtained with GMOS on the Gemini North
telescope in the months of February and April 2003.  GMOS masks for
three fields were designed, but only the central field was observed
within the time allocation. Seeing was typically $\sim$0.9 arcsec.

The GMOS CCDs consist of 3 abutted 2048$\times$4608 EEV chips with a
plate scale of 0.0727 arcsec/pixel (un-binned). For our set-up, we
binned 4$\times$ in the dispersion direction, yielding 1.84\AA\ per
binned pixel, giving a final resolution of FWHM $\sim$4\AA. The
dispersion runs across the detectors in the GMOS instrument, resulting
in two $\sim$20\AA\ gaps in the dispersion direction of the spectra.
The B600\_G5303 grating was used, with central wavelengths for
successive exposures set alternately to 5000\AA\ and 5025\AA\ (in
order to obtain full wavelength information across the gaps in the
GMOS detectors).  The effective wavelength range of each slit-let is a
function of its position on the mask, but typically covered
3800\AA\--6660\AA. A slit width of 1.0 arcsec was chosen to match the
seeing and the minimum slit length was 6 arcsec, a compromise between
maximising the number of slit-lets and allowing for reliable
sky-subtraction.  Exposures of 20$\times$1800s were taken, yielding a
total of 10 hours on-source integration time.  Bias frames, dome
flat-fields and Copper-Argon (CuAr) arc exposures were taken as part
of the Gemini baseline calibrations.

These data were reduced using the Gemini/GMOS packages in IRAF and a
number of custom made scripts. From the CuAr arcs, wavelength
solutions with typical residuals of 0.1\AA\ were achieved. Objects and
sky regions in the object spectra were manually identified in
cross-sections of the two-dimensional images. The spectra were then
extracted by tracing them in the dispersion direction.  After some
experimentation, optimal (variance) extraction was found to yield the
best results since our data are over-sampled on the detector. In some
cases, objects were too faint to trace individually and we therefore
co-added several 2-d images, taken adjacent in time, to act as a
reference for the extractions. We verified beforehand that flexure was
minimal between the reference images.  Finally, the extracted spectra
were median combined and weighted by their fluxes with cosmic ray
rejection.

\begin{figure}
\centerline{\psfig{figure=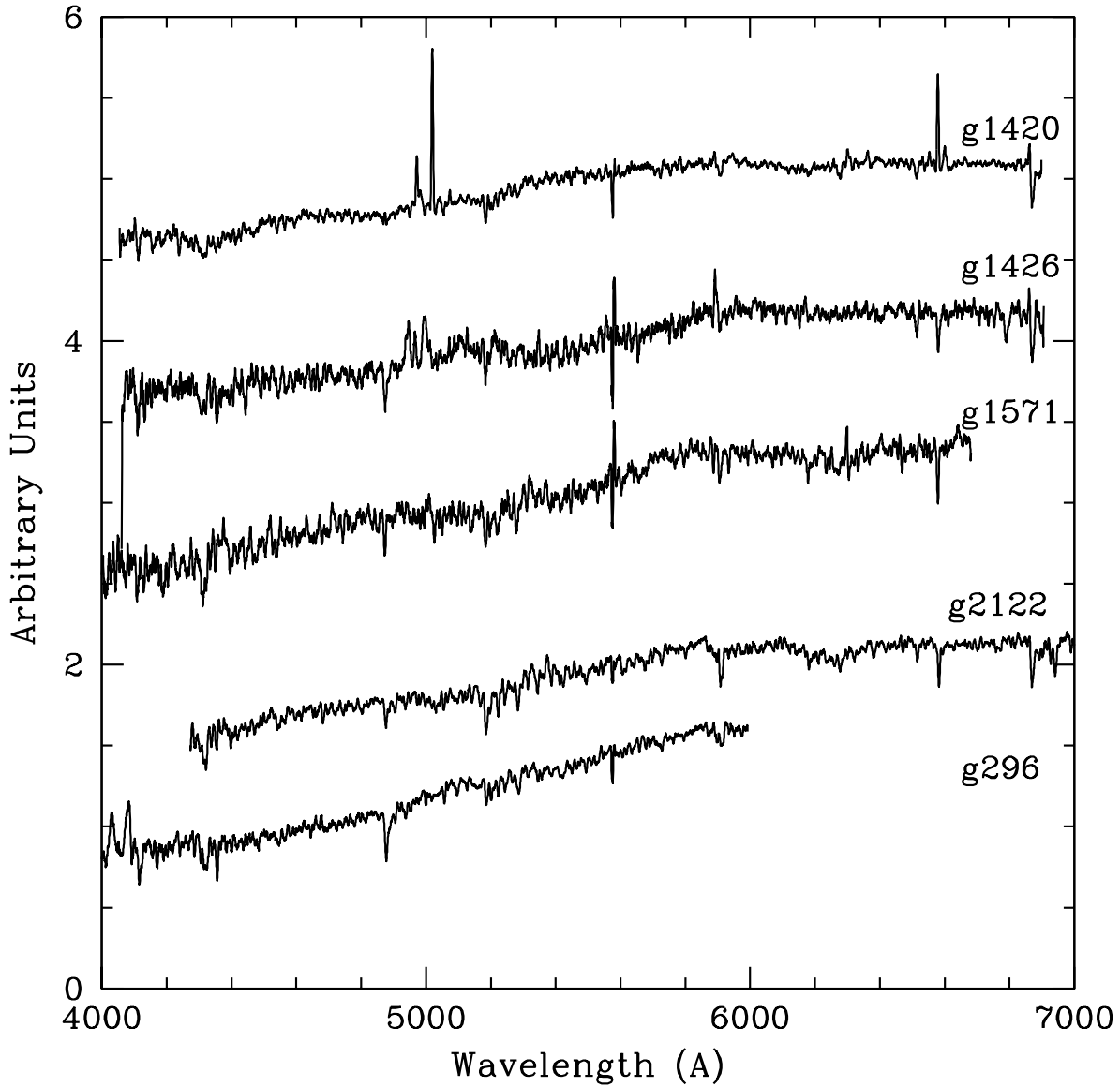,width=1.0\textwidth,angle=0}}
\caption{Normalised GC spectra which have been offset by one unit.  These spectra have not been de-redshifted. These sample spectra show the wavelength range that the majority of our spectra cover and display the range of S/N and metallicity present.  The spectrum of g1426 shows emission features around 5000 \AA\ and 5890\AA\ of unknown origin.  The PN hosting GC, g1420, is also plotted and the emission due to 4959 and 5007 \AA\ [OIII] lines can be seen redshifted to 4971 and 5019 \AA.}
\label{spec}
\end{figure}

In the absence of any velocity standard stars, the recession
velocities were measured by using four Bruzual \& Charlot (2003) model
stellar energy distributions (SEDs) for old ages with metallicities
[Fe/H] = --2.25, --1.64, --0.64 and --0.33. The task FXCOR in IRAF was
used and the average (weighted by the cross-correlation peak-height)
was taken.  Objects with recession velocities in the range 900$\pm$400
km/s are potentially associated with NGC~3379.  These are presented in
Table \ref{tabobs} and also used in our kinematic analysis in Section
\ref{sec_kin}. There was one foreground star and one background galaxy
out of the 24 spectra obtained.  Our low contamination rate of 9\% is
due to good imaging and colour selection.  We note that the spectrum
of g1420 shows emission lines, most notable 4959 and 5007 \AA\ [OIII]
which we assume are due to a planetary nebula (see Fig. \ref{spec}).
Minniti \& Rejkuba (2002) report a similar discovery for a GC in
NGC~5128 and provide a discussion of the implications.  Another GC,
g1426, shows unexplained emission features around $\sim$5000 \AA.

A first-order flux calibration was carried out by normalising the
spectra with a low-order polynomial. In order to measure Lick indices,
we convolved our spectra with a wavelength-dependent Gaussian kernel
to match the resolution of the Lick/IDS system (see Beasley \etal
2004b). Lick indices (Trager \etal 1998) were measured from our
normalised spectra.  Due to the variable wavelength ranges in these
spectra, the same set of indices could not be measured for all
spectra. However, all covered a wavelength range of
4500--5500~\AA. Uncertainties in the indices were derived from the
photon noise in the unfluxed spectra.  No Lick standard stars were
observed so we therefore cannot fully calibrate the GCs onto the Lick
system.  Consequently there are some systematic differences between
some of the measured indices and those used in the SSP models.  These
issues are discussed further in Sections \ref{sec_age} and
\ref{sec_disc}. Measured line indices and uncertainties are presented
in Tables \ref{tabindex1} and \ref{tabindex2}.

The final spectra have S/N = 18--58 \AA$^{-1}$ at 5000 \AA, giving
errors in the H$\beta$ index of 0.13--0.44 \AA\ (we note that g1566
has an H$\beta$ error of $\pm$0.065 \AA\ and S/N=118, however this is
not representative as this GC is $\sim$1.5 mags brighter than any
other). 

\begin{table*} 
\begin{center}
\caption{\scriptsize Confirmed Globular Clusters around NGC~3379.  Cluster ID, coordinates, {\it g} magnitude, {\it g--r} and {\it g--i} colours are from our GMOS imaging and are instrumental magnitudes only.  Heliocentric velocities are from the spectra presented in this work.}
\renewcommand{\arraystretch}{1.5} 
\begin{tabular}{lcccccc} 
\hline 
\hline 
ID     & R.A.   & Dec.       &   {\it g} & {\it g--r} & {\it g--i}  & V$_{helio}$\\ 
       & (J2000) & (J2000)   &  (mag) & (mag) &  (mag) & (km/s)\\ 
\hline
g1566 &  10:47:50.5 & 12:34:37.2  &  19.83 &  0.65 &   0.96 &  1071$\pm$18 \\
g296  &  10:48:02.2 & 12:35:57.6  &  21.40 &  0.62 &   0.88 &  1022$\pm$52 \\
g1940 &  10:47:48.2 & 12:35:45.3  &  21.42 &  0.74 &   0.90 &  875$\pm$25 \\
g1550 &  10:47:53.6 & 12:34:54:6  &  21.47 &  0.71 &   1.09 &  1100$\pm$31  \\
g2122 &  10:47:47.7 & 12:34:14.8  &  21.49 &  0.74 &   0.93 &  922$\pm$32 \\
g1078 &  10:47:54.2 & 12:36:31.9  &  21.54 &  0.70 &   1.00 &  1000$\pm$24 \\
g1420 &  10:47:50.7 & 12:35:32.3  &  21.54 &  0.62 &   0.90 &  764$\pm$38 \\
g1540 &  10:47:53.5 & 12:35:04.9  &  21.68 &  0.58 &   0.83 &  790$\pm$41 \\
g1617 &  10:47:50.4 & 12:33:47.8  &  21.88 &  0.62 &   0.96 &  650$\pm$37 \\
g1588 &  10:47:49.7 & 12:34:10.2  &  22.32 &  0.60 &   1.10 &  546$\pm$45 \\
g114  &  10:48:04.9 & 12:35:38.5  &  22.35 &  0.61 &   0.90 &  861$\pm$45 \\
g1426 &  10:47:50.6 & 12:35:19.2  &  22.37 &  0.61 &   0.87 &  797$\pm$71 \\
g1610 &  10:47:58.9 & 12:33:54.5  &  22.39 &  0.55 &   0.79 &  957$\pm$61 \\
g1571 &  10:47:49.7 & 12:34:32.8  &  22.47 &  0.72 &   1.04 &  744$\pm$28  \\
g1519 &  10:47:49.6 & 12:35:27.0  &  22.49 &  0.73 &   1.16 &  623$\pm$42 \\
g1934 &  10:47:47.9 & 12:36:23.4  &  22.72 &  0.64 &   0.76 &  1004$\pm$65 \\
g1580 &  10:47:54.2 & 12:34:21.6  &  22.84 &  0.71 &   1.09 &  1066$\pm$32 \\
g219  &  10:48:01.8 & 12:37:22.0  &  23.04 &  0.57 &   0.85 &  868$\pm$95 \\
g1893 &  10:47:45.1 & 12:37:53.0  &  23.26 &  0.86 &   1.21 &  804$\pm$43 \\
g1630 &  10:47:50.5 & 12:36:11.8  &  23.65 &  0.77 &   1.21 &  1115$\pm$61 \\
g1595 &  10:47:52.9 & 12:34:04.4  &  23.69 &  0.67 &   1.00 &  970$\pm$70 \\
g230  &  10:48:01.2 & 12:37:06.9  &  23.93 &  0.80 &   1.28 &  1022$\pm$90 \\
\hline 
\label{tabobs}
\end{tabular} 
\end{center} 
\end{table*}

\begin{table*} 
\begin{center}
\caption{\scriptsize Globular Cluster indices $\lambda <$ 4600 \AA.  Central index values (first line) and errors (second line).  Indices in brackets are removed during the fitting process.   Missing values are due to limited wavelength coverage.}
\renewcommand{\arraystretch}{1.5}
\begin{tabular}{lcccccccccccc}
\hline 
\hline 
ID&H$\delta_{A}$&H$\delta_{F}$&CN$_{1}$&CN$_{2}$&Ca4227&G band&H$\gamma_{A}$&H$\gamma_{F}$&Fe4383&Ca4455& Fe4531 \\
     & (\AA) & (\AA) & (mag) & (mag) & (\AA) & (\AA) & (\AA) & (\AA) &  (\AA) & (\AA) & (\AA) \\
\hline
g1566  &  ...... &  ...... & ....... & ...... &  0.493 &  3.106 & -0.962 &  0.896 &  2.365 &  0.591 &   2.008 \\
      &  ...... &  ...... & ....... & ...... &  0.069 &  0.117 &  0.119 &  0.073 &  0.167 &  0.084 &   0.124 \\
g296   &  (4.054)&  (3.555)& (0.002) & (0.038)&  0.481 & 2.887  &-0.409  &  1.019 &  0.707 &  0.794 &   0.856 \\
      &  (0.253)&  (0.166)& (0.008) & (0.009)&  0.138 & 0.237  & 0.238  &  0.148 &  0.353 &  0.173 &   0.266 \\
g1940  &  ...... &  ...... & ....... & ...... & ...... &  3.256 & -0.919 &  0.437 &  1.207 &  0.514 &   2.532 \\
      &  ...... &  ...... & ....... & ...... & ...... &  0.235 &  0.237 &  0.151 &  0.342 &  0.169 &   0.243 \\
g1550  &  ...... &   1.302 & (0.057) & (0.106)&  0.681 &  4.005 &(-4.947)&(-1.474)&  3.992 & (0.467)&   3.044 \\
      &  ...... &   0.205 & (0.008) & (0.010)&  0.151 &  0.246 & (0.273)& (0.173)&  0.342 & (0.179)&   0.255 \\
g2122  &  ...... &  ...... & ....... & ...... & ...... &  5.555 & -5.405 & -1.191 &  4.199 & (0.866)&   3.108 \\
      &  ...... &  ...... & ....... & ...... & ...... &  0.251 &  0.295 &  0.184 &  0.363 & (0.185)&   0.266 \\
g1078  &  0.865  &   1.694 & (0.010) & (0.043)&  0.588 &  3.866 & -1.420 &  0.486 & (1.280)& (0.285)&   2.309 \\
      &  0.291  &   0.195 & (0.008) & (0.009)&  0.146 &  0.241 &  0.249 &  0.157 & (0.354)& (0.177)&   0.250 \\
g1420  &  ...... &  (1.278)& (0.019) & (0.049)& (1.089)&  2.719 & -0.457 & (0.541)&  0.763 &  0.504 &  (0.770)\\
      &  ...... &  (0.193)& (0.008) & (0.009)& (0.139)&  0.254 &  0.249 & (0.159)&  0.366 &  0.178 &  (0.266)\\
g1540  &   3.926 &   2.973 &(-0.057) &(-0.053)&  0.439 &  2.091 &  0.367 & (1.201)&  0.877 &  0.477 &   0.846 \\
      &   0.270 &   0.185 & (0.008) & (0.009)&  0.149 &  0.255 &  0.245 & (0.155)&  0.369 &  0.182 &   0.277 \\
g1617  &  ...... &  (0.724)&(-0.021) & (0.017)& (0.191)&  3.743 & -1.765 &  0.525 &  2.259 &  0.447 &   1.943 \\
      &  ...... &  (0.240)& (0.009) & (0.011)& (0.177)&  0.292 &  0.302 &  0.188 &  0.415 &  0.209 &   0.305 \\
g1588  &  ...... &  ...... & ....... & ...... &  1.192 &  4.292 &(-4.894)& -0.283 &  3.098 &  0.837 &   3.327 \\
      &  ...... &  ...... & ....... & ...... &  0.213 &  0.350 & (0.386)&  0.234 &  0.492 &  0.249 &   0.356 \\
g114   &  (0.829)&   2.917 &(-0.079) &(-0.079)&(-0.314)&  2.070 &  0.402 &  1.935 &  0.551 &  0.648 & (-0.256)\\
      &  (0.432)&   0.270 & (0.012) & (0.014)& (0.221)&  0.382 &  0.357 &  0.218 &  0.534 &  0.259 &  (0.398)\\
g1426  &  ...... &   2.368 &(-0.037) &(-0.008)&(-0.208)&  4.122 &  0.267 &  0.525 &  0.860 &  0.051 &   1.447 \\
      &  ...... &   0.268 & (0.011) & (0.013)& (0.218)&  0.342 &  0.351 &  0.230 &  0.521 &  0.259 &   0.389 \\
g1610  &  (4.667)&   3.479 &(-0.234) &(-0.232)&  0.175 &(-2.291)& -1.207 &  0.580 &  1.182 &  0.678 &   1.332 \\
      &  (0.380)&   0.261 & (0.012) & (0.014)&  0.215 & (0.433)&  0.363 &  0.229 &  0.520 &  0.250 &   0.390 \\  
g1571  &  (3.298)&  (1.842)& (0.001) & (0.042)&  0.859 & (6.656)&(-8.418)& -1.715 &  6.113 &  2.101 &   2.561 \\
      &  (0.439)&  (0.315)& (0.013) & (0.015)&  0.230 & (0.350)& (0.448)&  0.268 &  0.498 &  0.260 &   0.397 \\
g1519  &  ...... &  ...... & ....... & ...... &  1.238 &  3.570 & -2.920 &  0.183 &  3.012 &  0.952 &   2.863 \\
      &  ...... &  ...... & ....... & ...... &  0.205 &  0.369 &  0.389 &  0.239 &  0.515 &  0.252 &   0.366 \\
g1934  &  ...... &  ...... & ....... & ...... & ...... &  0.262 &  1.058 &  1.875 & -0.069 &  0.517 &  -0.171 \\
      &  ...... &  ...... & ....... & ...... & ...... &  0.427 &  0.384 &  0.238 &  0.606 &  0.300 &   0.462 \\
g1580  &  -1.848 & (-1.852)& (0.177) & (0.129)&(-1.926)&  3.971 & -5.028 & -0.634 &  5.124 &  2.185 &  (0.530)\\
      &   0.659 &  (0.475)& (0.017) & (0.020)& (0.367)&  0.487 &  0.526 &  0.323 &  0.631 &  0.308 &  (0.494)\\
g219   & (-4.486)&  (0.019)& (0.037) & (0.188)& (2.267)& -1.395 &  4.035 & (1.342)&  0.040 & -0.334 &  -0.274 \\
      &  (0.665)&  (0.422)& (0.016) & (0.018)& (0.253)&  0.613 &  0.473 & (0.325)&  0.767 &  0.385 &   0.559 \\
g1893  &  ...... &  ...... & ......  & ...... & ...... & ...... & ...... & ...... & ...... & ...... &   2.245 \\
      &  ...... &  ...... & ......  & ...... & ...... & ...... & ...... & ...... & ...... & ...... &   0.602 \\
g1630  &  ...... & (-5.925)& (0.150) &(0.121) & (3.212)&  5.462 & -8.396 & -1.286 &  6.183 &(-1.965)&  (5.771)\\
      &  ...... &  (0.927)& (0.025) &(0.029) & (0.339)&  0.650 &  0.851 &  0.487 &  0.960 & (0.555)&  (0.615)\\
g1595  & (-0.313)& (-3.507)&(-0.182) &(-0.060)& -0.684 & -2.792 &  2.502 & (4.899)& (7.957)& (3.874)&  (5.532)\\
      &  (1.160)&  (0.986)& (0.030) & (0.034)&  0.575 &  1.038 &  0.761 & (0.414)& (0.961)& (0.498)&  (0.773)\\
g230   & (11.407)&   2.689 &(-0.644) &(-0.585)&(-3.332)&(-4.732)&  2.696 &(-3.572)&(-3.244)& -0.978 &  (6.999)\\
      &  (0.813)&   0.766 & (0.028) & (0.031)& (0.645)& (1.047)&  0.726 & (0.612)& (1.213)&  0.556 &  (0.671)\\
\hline 
\label{tabindex1}
\end{tabular} 
\end{center} 
\end{table*} 

\begin{table*} 
\begin{center} 
\caption{\scriptsize Globular Cluster indices $\lambda >$4600 \AA.  Central index values (first line) and errors (second line).  Indices in brackets are removed during the fitting process.  Missing values are due to limited wavelength coverage.}
\renewcommand{\arraystretch}{1.5}
\begin{tabular}{lccccccccccc} 
\hline 
\hline 
ID  &  C4668 & H$\beta$ & Fe5015 & Mg$_{1}$ & Mg$_{2}$ &  Mgb   & Fe5270 & Fe5335 & Fe5406 & Fe5709 & Fe5782 \\
   & (\AA)  & (\AA)    & (\AA)  & (mag)    & (mag)    & (\AA)  & (\AA)  & (\AA)  & (\AA)  & (\AA)  & (\AA) \\
\hline 
g1566 & (1.474)&  1.994 &  2.692 &  (0.039) &  (0.110) &  1.895 &  1.668 & (1.606)&  0.832 &  0.467 &  0.242 \\
     & (0.180)&  0.065 &  0.139 &  (0.001) &  (0.002) &  0.064 &  0.070 & (0.080)&  0.059 &  0.044 &  0.042  \\
g296  &  0.339 & (2.679)&  1.676 & (-0.039) &  (0.028) & (1.027)&  1.194 &  0.819 & (0.778)& ...... & ...... \\
     &  0.382 & (0.132)&  0.283 &  (0.003) &  (0.003) & (0.130)&  0.142 &  0.163 & (0.119)& ...... & ......  \\
g1940 &  1.193 &  2.153 &  2.816 &  (0.008) &  (0.093) &  2.045 &  1.745 &  1.425 & (1.250)& (0.772)&  0.178 \\
     &  0.362 &  0.131 &  0.279 &  (0.003) &  (0.003) &  0.128 &  0.142 &  0.161 & (0.117)& (0.087)&  0.084  \\
g1550 & (3.430)& (1.100)&  3.697 &  (0.085) &  (0.198) &  2.810 &  2.206 &  1.687 &  1.354 &  0.727 &  0.391 \\
     & (0.369)& (0.139)&  0.287 &  (0.003) &  (0.003) &  0.133 &  0.143 &  0.163 &  0.118 &  0.089 &  0.085  \\
g2122 &  2.411 &  1.670 &  3.940 &  (0.065) &  (0.200) &  2.953 &  2.387 & (2.850)&  1.546 &  0.590 & (0.346)\\
     &  0.394 &  0.144 &  0.304 &  (0.003) &  (0.004) &  0.139 &  0.150 & (0.165)&  0.123 &  0.093 & (0.089) \\
g1078 & (3.035)&  2.147 & (4.188)&  (0.025) &  (0.112) &  2.118 &  1.495 &  1.113 &  1.115 &  0.685 &  0.314 \\
     & (0.366)&  0.134 & (0.280)&  (0.003) &  (0.003) &  0.130 &  0.144 &  0.165 &  0.119 &  0.087 &  0.084  \\
g1420 &  1.273 & (0.938)&(-0.436)&  (0.038) &  (0.115) &  1.418 &  1.153 &  1.115 &  0.694 &  0.250 &  0.302 \\
     &  0.380 & (0.143)& (0.296)&  (0.003) &  (0.003) &  0.136 &  0.147 &  0.166 &  0.121 &  0.093 &  0.088  \\
g1540 &(-1.675)&  2.934 &(-0.446)&  (0.016) &  (0.056) &  1.441 &  0.717 &  0.898 & (0.973)&(-0.433)& -0.156 \\
     & (0.414)&  0.139 & (0.325)&  (0.003) &  (0.004) &  0.144 &  0.163 &  0.184 & (0.133)& (0.101)&  0.095  \\
g1617 &  1.517 & (1.131)&  2.076 & (-0.032) &  (0.076) &  2.534 &  1.268 &  0.814 &  0.354 &  0.288 &  0.163 \\
     &  0.448 & (0.169)&  0.352 &  (0.003) &  (0.004) &  0.155 &  0.176 &  0.203 &  0.148 &  0.112 &  0.106  \\
g1588 &(-0.540)&  2.255 &(-0.892)&  (0.081) &  (0.167) &  2.413 & (3.422)&  1.600 &(-0.020)&  0.700 & (0.002)\\
     & (0.546)&  0.187 & (0.418)&  (0.004) &  (0.005) &  0.185 & (0.195)&  0.226 & (0.175)&  0.121 & (0.118) \\
g114  & (2.781)&  2.327 &  2.621 &  (0.001) &  (0.051) &  1.230 &  1.687 &  1.310 &  0.706 & ...... & ...... \\
     & (0.558)&  0.197 &  0.426 &  (0.004) &  (0.005) &  0.196 &  0.212 &  0.245 &  0.180 & ...... & ......  \\
g1426 &  0.957 &  2.962 &(-0.001)& (-0.067) & (-0.005) &  2.373 &(-1.293)& -0.045 &  0.920 &(-0.364)&  0.161 \\
     &  0.570 &  0.196 & (0.433)&  (0.004) &  (0.005) &  0.188 & (0.240)&  0.267 &  0.188 & (0.143)&  0.131  \\
g1610 &  0.796 &  2.451 &  3.175 &  (0.024) &  (0.075) & (0.797)&  0.906 &  0.900 &  0.419 & ...... & ...... \\
     &  0.576 &  0.207 &  0.441 &  (0.004) &  (0.005) & (0.208)&  0.222 &  0.252 &  0.189 & ...... & ......  \\
g1571 &  1.771 &  0.976 & (0.318)&  (0.085) &  (0.157) &  2.159 &  2.677 &  2.220 & (0.742)&(-0.124)&  0.705 \\
     &  0.578 &  0.214 & (0.464)&  (0.004) &  (0.005) &  0.207 &  0.221 &  0.249 & (0.192)& (0.138)&  0.126  \\
g1519 &  0.895 & (0.468)& (6.229)& (-0.003) &  (0.189) & (4.168)& (3.613)&  1.874 &(-0.074)& (1.240)&  0.362 \\
     &  0.549 & (0.213)& (0.406)&  (0.004) &  (0.005) & (0.180)& (0.190)&  0.226 & (0.173)& (0.129)&  0.126  \\
g1934 &  0.258 &  3.209 &(-2.122)& (-0.023) & (-0.052) &(-3.180)&  1.031 &  0.496 & (0.961)&  0.157 & -0.198 \\
     &  0.662 &  0.231 & (0.533)&  (0.005) &  (0.006) & (0.260)&  0.264 &  0.301 & (0.216)&  0.168 &  0.167  \\
g1580 &  2.831 & (3.334)& (7.071)&  (0.042) &  (0.166) & (3.601)&  2.713 & (0.901)& (0.421)&  0.856 &(-0.083)\\
     &  0.696 & (0.246)& (0.518)&  (0.005) &  (0.006) & (0.236)&  0.260 & (0.306)& (0.225)&  0.163 & (0.161) \\
g219  &(-2.680)&  2.595 &  0.108 & (-0.034) &  (0.043) &  1.345 &  0.930 & -0.152 &  0.715 & ...... & ...... \\
     & (0.809)&  0.274 &  0.614 &  (0.006) &  (0.007) &  0.271 &  0.299 &  0.346 &  0.246 & ...... & ......  \\
g1893 &  7.066 & (3.030)&  7.438 &  (0.111) &  (0.250) &  5.102 & (4.318)&  2.351 &  2.286 &  1.414 & (0.316)\\
     &  0.818 & (0.305)&  0.615 &  (0.006) &  (0.007) &  0.273 & (0.293)&  0.345 &  0.254 &  0.191 & (0.182) \\ 
g1630 & (1.358)&  2.091 & (0.673)&  (0.145) &  (0.238) &  2.505 &  2.254 &  2.013 &(-0.329)&  1.568 &  1.053 \\
     & (0.979)&  0.354 & (0.827)&  (0.007) &  (0.009) &  0.351 &  0.366 &  0.417 & (0.322)&  0.219 &  0.212  \\
g1595 &(12.186)&  2.914 &  0.448 &  (0.043) &  (0.066) & (3.838)& -0.352 &(-2.982)& -0.690 & (1.311)& -0.223 \\
     & (1.024)&  0.396 &  0.897 &  (0.009) &  (0.010) & (0.366)&  0.451 & (0.543)&  0.386 & (0.268)&  0.268  \\
g230  & (9.738)& (0.763)&  0.969 & (-0.113) &  (0.067) &  0.724 & (4.403)&  1.941 &(-1.199)& ...... & ...... \\
     & (1.046)& (0.442)&  0.856 &  (0.008) &  (0.010) &  0.405 & (0.395)&  0.466 & (0.388)& ...... & ......  \\
\hline 
\label{tabindex2}
\end{tabular} 
\end{center} 
\end{table*}

\section{Ages, Metallicities and $\alpha$-Element Abundance Ratios}\label{sec_age}

In this section we describe how we derive ages, metallicities and
$\alpha$-element abundances.  The resulting values are listed in Table
\ref{tabagez}.

We apply the $\chi^{2}$ multi-index fitting technique of Proctor \&
Sansom (2002) for this analysis.  This method involves the comparison
of the measured Lick indices with SSP models (its application to
extra-galactic GCs is described fully in Pierce \etal 2005).  The SSP
models of Thomas, Maraston \& Korn (2004; hereafter TMK04) were chosen
because they present the only set of models that include the effect of
$\alpha$ abundance ratios on the Balmer lines.

We compare the measured Lick indices to the TMK04 SSPs and obtain a
minimum $\chi^2$ fit. This fit is obtained using all the indices
measured. Simultaneously a set of $\chi^2$ minimisation fits are found
with each of the indices omitted.  From this set, we select the fit
with the lowest total $\chi^2$, remove the necessary index and repeat
until a stable fit is achieved with no highly aberrant ($> 3 \sigma$)
indices remaining. All GCs had some indices that were significant
outliers to the fit and therefore removed during this process.  For
g1426 none of the indices potentially affected by the PN style
emission lines are included in the final fit.

The molecular band indices Mg$_1$ and Mg$_2$ are systematically offset
due to poor flux calibration and were excluded for all GCs (see
Proctor \etal 2005). Similar to other GC studies (e.g. Beasley \etal
2004a) we find the CN indices to be enhanced relative to the models
and therefore they were also removed.

\begin{table*} 
\begin{center} 
\caption{\scriptsize Derived Globular Cluster Properties.  Age, [Fe/H], [E/Fe] and [Z/H] are derived from the $\chi^2$ minimisation process, with errors derived by a Monte Carlo style method.  [Fe/H]$_{BH}$ is derived according to the method of Brodie \& Huchra (1990) from a reduced sample of indices.}
\renewcommand{\arraystretch}{1.5} 
\begin{tabular}{lccccc} 
\hline 
\hline 
ID     & Age   & [Fe/H]    & [E/Fe]    & [Z/H] & [Fe/H]$_{BH}$ \\ 
       &(Gyr)  & (dex)     & (dex)     & (dex) & (dex)        \\ 
\hline 
g1566 &  12.6$\pm$1.1 & --1.17$\pm$0.05 &   0.15$\pm$0.05  & --1.03$\pm$0.03  &  --1.15$\pm$0.06 \\
g296  &  11.2$\pm$1.5 & --1.82$\pm$0.11 &   0.71$\pm$0.10  & --1.15$\pm$0.06  &  --1.84$\pm$0.26 \\
g1940 &  11.9$\pm$2.1 & --1.26$\pm$0.11 &   0.30$\pm$0.10  & --0.98$\pm$0.06  &  --1.31$\pm$0.12 \\
g1550 &  15.0$\pm$4.8 & --0.69$\pm$0.11 &   0.15$\pm$0.07  & --0.55$\pm$0.11  &  --0.52$\pm$0.18 \\
g2122 &  15.0$\pm$2.0 & --0.47$\pm$0.07 &   0.15$\pm$0.06  & --0.33$\pm$0.04  &  --0.45$\pm$0.10 \\ 
g1078 &  10.0$\pm$1.7 & --1.05$\pm$0.09 &   0.24$\pm$0.09  & --0.83$\pm$0.06  &  --1.17$\pm$0.06 \\
g1420 &  10.0$\pm$1.4 & --1.40$\pm$0.11 &   0.21$\pm$0.09  & --1.20$\pm$0.06  &  --1.23$\pm$0.10 \\
g1540 &   9.4$\pm$1.7 & --2.09$\pm$0.18 &   0.68$\pm$0.17  & --1.45$\pm$0.09  &  --1.64$\pm$0.04 \\
g1617 &  11.2$\pm$1.9 & --1.52$\pm$0.12 &   0.68$\pm$0.10  & --0.88$\pm$0.08  &  --1.59$\pm$0.28 \\ 
g1588 &   8.4$\pm$2.9 & --0.57$\pm$0.14 &   0.15$\pm$0.09  & --0.43$\pm$0.13  &  --0.33$\pm$0.16 \\
g114  &  15.0$\pm$6.5 & --1.14$\pm$0.16 & --0.06$\pm$0.15  & --1.20$\pm$0.14  &  --1.60$\pm$0.17  \\
g1426 &   7.5$\pm$2.5 & --1.58$\pm$0.13 &   0.80$\pm$0.07  & --0.83$\pm$0.11  &  --2.37$\pm$0.44  \\
g1610 &  10.6$\pm$2.1 & --1.62$\pm$0.13 &   0.71$\pm$0.12  & --0.95$\pm$0.09  &  --1.98$\pm$0.40 \\
g1571 &  15.0$\pm$3.6 & --0.10$\pm$0.08 & --0.27$\pm$0.05  & --0.35$\pm$0.08  &  --0.29$\pm$0.21  \\
g1519 &  12.6$\pm$2.9 & --0.72$\pm$0.13 & --0.03$\pm$0.10  & --0.75$\pm$0.09  &  --0.82$\pm$0.37  \\
g1934 &  10.0$\pm$2.3 & --2.31$\pm$0.31 &   0.65$\pm$0.29  & --1.70$\pm$0.15  &  --2.27$\pm$0.21 \\
g1580 &   9.4$\pm$2.5 & --0.24$\pm$0.13 &   0.09$\pm$0.09  & --0.15$\pm$0.09  &  --0.73$\pm$0.12  \\
g219  &  12.6$\pm$1.8 & --2.72$\pm$0.37 &   0.50$\pm$0.35  & --2.25$\pm$0.13  &  --2.15$\pm$0.25 \\
g1893 &  11.2$\pm$5.9 &   0.12$\pm$0.16 &   0.24$\pm$0.09  &   0.35$\pm$0.13  &    0.44$\pm$0.05 \\
g1630 &  11.2$\pm$5.4 &   0.10$\pm$0.19 & --0.24$\pm$0.08  & --0.13$\pm$0.19  &    0.20$\pm$0.35 \\
g1595 &   8.9$\pm$3.2 & --3.00$\pm$0.49 &   0.80$\pm$0.41  & --2.25$\pm$0.31  &  --2.08$\pm$0.42 \\
g230  &  15.0$\pm$5.9 & --1.52$\pm$0.30 & --0.30$\pm$0.34  & --1.80$\pm$0.23  &  --2.34$\pm$0.94  \\
\hline 
\label{tabagez}
\end{tabular} 
\end{center} 
\end{table*} 

\begin{figure}
\centerline{\psfig{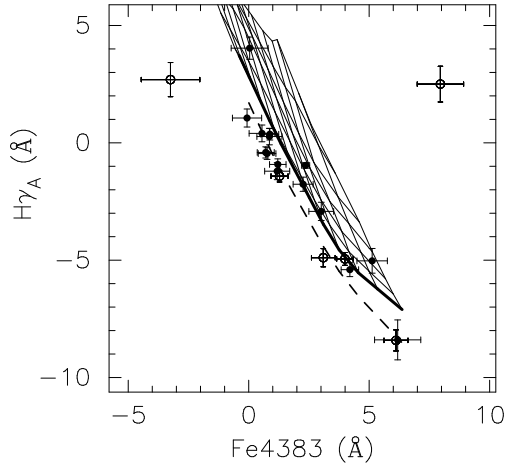}}
\caption{The indices H$\gamma_A$ vs Fe4383 are plotted against age-metallicity grids from TMK04. We show the grid for [E/Fe]=+0.3, with metallicity in 0.25 dex steps from from --2.25 (left) to +0.5 (right) and ages of 1,2,3,5,8,12 and 15 Gyrs (top to bottom), the bold line is 15 Gyrs and the dashed line is the red horizontal branch 15 Gyr isochrone from Thomas, Maraston \& Bender (2003). Filled circles are shown for GCs where both indices are included in the fits, if either index is excluded then an open circle is plotted. A large number of the GCs fall near the maximum age line for the Thomas, Maraston \& Bender (2003) red horizontal branch SSPs.}
\label{indexplot}
\end{figure}

An index-index plot of H$\gamma_A$ vs Fe4383 is shown in
Fig. \ref{indexplot}.  This pairing of indices was chosen as the
combination of an age and a metallicity sensitive index that had the
largest number of reliable points. This shows the majority of GCs lie
near the 15 Gyr age line.  Horizontal branch morphology is the
apparent source of the offset to the left of the TMK04 15 Gyr age line
seen in the majority of the low metallicity GCs (see Pierce \etal 2005
for a brief discussion on the effect of horizontal branch morphology
on GC $\chi^2$ fitting).  TMK04 models in this regime posses a
predominantly blue horizontal branch whereas the red horizontal branch
models from Thomas, Maraston \& Bender (2003) pass directly through
these points. There are a couple of clear outliers for which one or
both index values have been excluded from the fitting process.
However to derive values we use the results from the $\chi^2$ fitting
process.

\begin{figure}
\centerline{\psfig{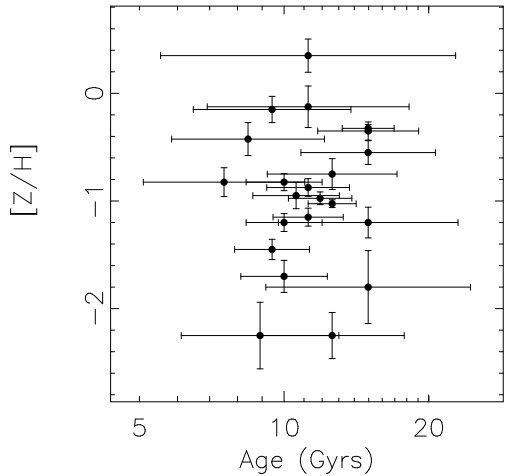}}
\caption{GC age-metallicity relation.  The plot shows that the observed GCs are consistent with an old age and a wide range of metallicities.}
\label{agefeplots}
\end{figure}

From our quantitative $\chi^2$ minimisation fits we find all of the
GCs to be consistent with an old age ($\ga$10 Gyrs) with a spread of
metallicities from [Fe/H] $<$ --2 to solar (Fig. \ref{agefeplots}).
This is consistent with Fig. \ref{indexplot} and indeed the fitting
procedure has identified clear outliers. The GC, g1426, with the
youngest fit age of 7.5 Gyrs, is the GC with unexplained emission
lines in its spectrum.  It also has [E/Fe]=+0.8 which is at the
maximum of the models.  While none of the clearly affected indices
were included in the fit, we suspect that the fit was still somewhat
influenced by the emission source.

\begin{figure}
\centerline{\psfig{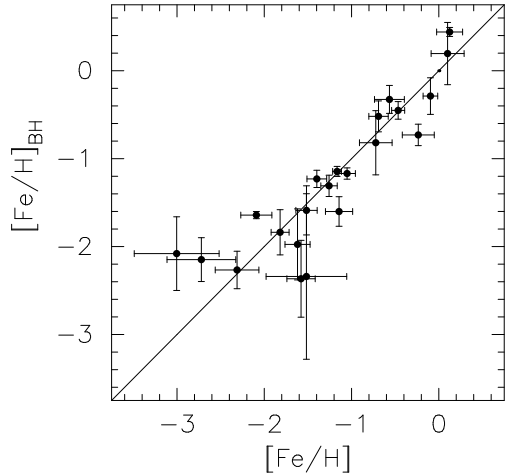}}
\caption{A comparison of Brodie-Huchra metallicity vs our derived metallicity.  The four points with poor agreement at low metallicity (below [Fe/H]$_{BH}$=~--2) all are at the limits of the SSP models. A gap in the metallicity distribution can be seen at [Fe/H]$\sim$~--1, which resembles that of the Milky Way GC system.}
\label{bhplot}
\end{figure}

A test of SSP-derived metallicities is to compare them with those
derived using the Brodie \& Huchra (1990; hereafter BH) method. We
were unable to use the full sample of metal-sensitive indices due to
the poor sensitivity at shorter wavelengths. We therefore measured the
G~band, Fe52, MgH and Mg$_2$ BH indices using the method outlined in
their paper. We define a Brodie-Huchra metallicity as the average of
the empirically calibrated metallicity from these 4 indices, where the
error quoted is the standard deviation. This metallicity is referred
to as [Fe/H]$_{BH}$ in Table \ref{tabagez}.  Fig. \ref{bhplot} shows
that the BH metallicities closely match those derived by $\chi^2$
fitting to SSPs. A gap in both metallicity measures occurs at
[Fe/H]$\sim$~--1. This is similar to that observed in the Milky Way GC
system (Harris 1996), which is also composed of an old GC population.
The outliers at low metallicity are at the limits of the SSP models
for either [Fe/H] or [E/Fe].

\begin{figure}
\centerline{\psfig{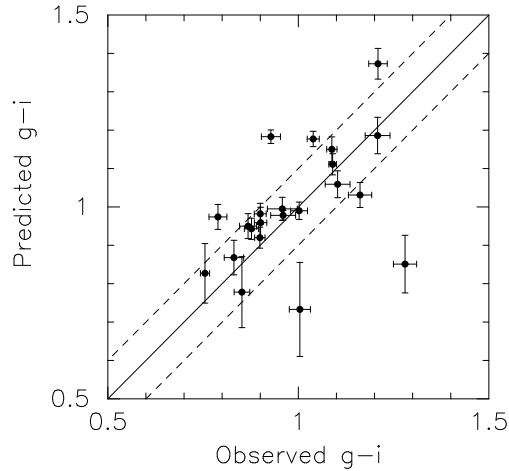}}
\caption{The predicted colour from  stellar population fits vs observed {\it g--i} colour. The outliers are GCs that are generally at the limits of the SSP models. The errors for both observed and predicted colours are lower limits and may be significantly underestimated.  The dashed lines show the RMS scatter of $\pm$0.1 mag around the one-to-one solid line. A Spearman Rank test gives a probability of correlation of 99.2\%.}
\label{colplots}
\end{figure}

We compare the observed {\it g--i} colours for the GCs with the
predicted {\it g--i} colours from the Bruzual \& Charlot (2003) SSPs,
with the ages and metallicities from TMK04 SSPs in Fig. \ref{colplots}
(thus there could be some systematic offsets due to model
differences). The use of Bruzual \& Charlot (2003) colour models is
necessary because the TMK04 SSPs do not include {\it g} and {\it i}
magnitudes. There are several outliers in the colour-colour plot,
these include the low metallicity GCs mentioned previously as well as
a few others that are at the SSP model maximum age. A Spearman Rank
test gives the probability of correlation as 99.2\%. Overall the figure
gives us confidence in the SSP-derived parameters.

\begin{figure}
\centerline{\psfig{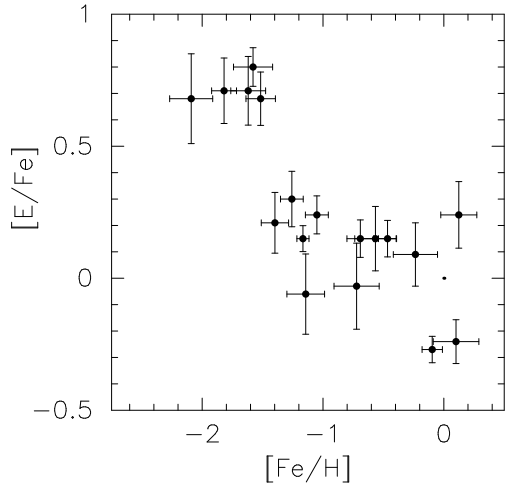}}
\caption{A plot of $\alpha$-element ratio vs metallicity. The four GCs with poorly determined metallicities and $\alpha$-element abundance ratios (i.e. [E/Fe] errors $>$0.3) are not shown. The plot indicates a correlation in the sense of increased enhancement with decreasing metallicity.}
\label{fealphaplots}
\end{figure}

\begin{figure}
\centerline{\psfig{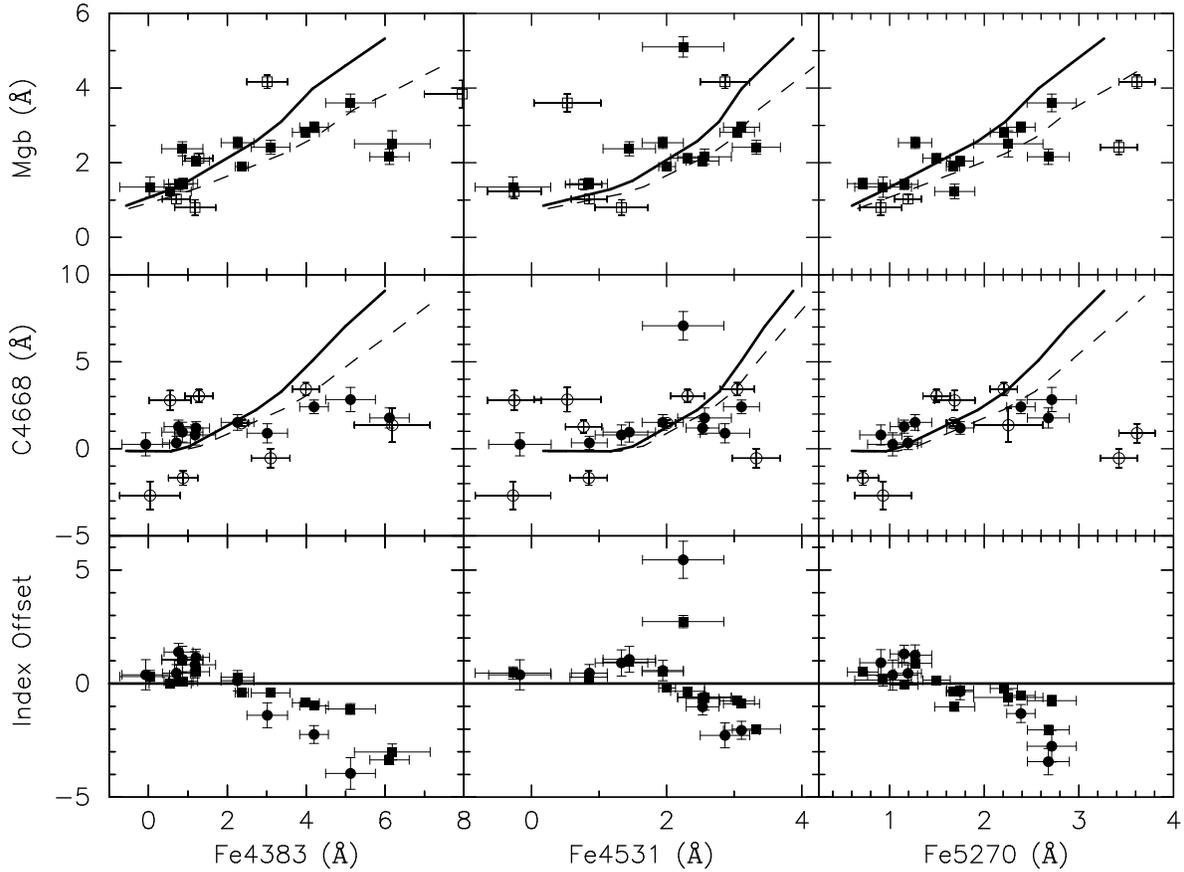}}
\caption{The upper row  show the $\alpha$-element
sensitive Mgb index vs three different Iron indices (Fe4383, Fe4531
and Fe5270).  The middle row shows C4668 vs the same Iron
indices.  All six have TMK04 12 Gyr isochrones plotted with
[E/Fe]=+0.3 as solid and [E/Fe]=+0.0 as dashed lines.  Filled symbols show
where both indices are included in the fit, if either of the indices
are not included in the final fit then open symbols are shown.  For
several plots one or two highly aberrant indices fall outside the
range plotted.  All six plots show that the $\alpha$-sensitive indices
are above the [E/Fe]=+0.3 dex model at low Iron sensitive index values
and are below for high Iron values.  The bottom row of plots show the
index offset relative to the [E/Fe]=+0.3 dex model line, for included
indices.  Square symbols are Mgb and the circles C4668. A strong trend
of decreasing $\alpha$-element ratio with increasing Iron abundance is
seen for both $\alpha$-element sensitive indices relative to the
[E/Fe]=+0.3 model expectation.}
\label{c46plots}
\end{figure}

TMK04 estimate $\alpha$-element abundance ratios using the parameter
[E/Fe].  A definition of [E/Fe] for the SSP models used can be found
in Thomas \etal (2004). Briefly, it includes $\alpha$-elements, such
as O, Ne, Mg, Si, S, Ar, Ca and Ti plus two non-$\alpha$-elements N
and Na. The [E/Fe] vs. [Fe/H] plot presented in
Fig. \ref{fealphaplots} shows decreasing $\alpha$-element abundance
ratio with increasing metallicity, including two solar metallicity GCs
with sub-solar $\alpha$-element abundance ratios.  The source of this
trend is confirmed by the index-index plots of Fig. \ref{c46plots}, in
which high-metallicity GCs generally have $\alpha$-sensitive indices
that are consistent with [E/Fe]$<$+0.3. The lower metallicity GCs have
$\alpha$-element sensitive indices that are predominantly stronger
than [E/Fe]=+0.3.  

\section{Globular Cluster Kinematics}\label{sec_kin}

In this section we use our measured velocities for 22 GCs (see Table
\ref{tabobs}) to estimate the radial velocity dispersion and
mass-to-light ratio for the halo of NGC~3379. Our sample is
supplemented by measurements of 14 GC velocities from Puzia \etal
(2004), who used the FORS instrument on the VLT.

\begin{figure}
\centerline{\psfig{figure=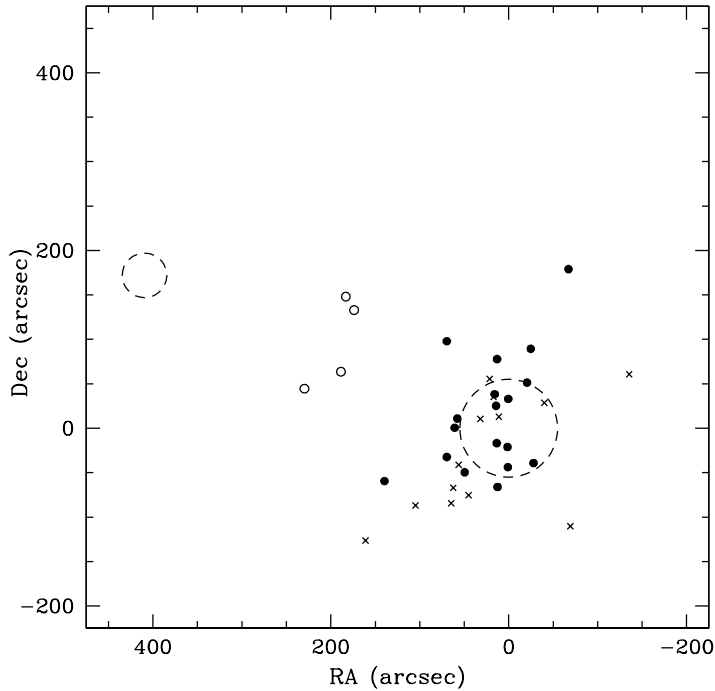,width=0.6\textwidth}}
\caption{Spatial positions of GCs. The circles are GMOS data presented here, crosses are VLT data of Puzia \etal (2004).  The two large circles show the effective radius of the galaxies NGC~3379 (lower right) and NGC~3384 (upper left) (de Vaucouleurs \etal 1991, Capaccioli \etal 1990).  The effective radius for NGC~3379 is 55$^{\prime\prime}$ (2.8 kpc). The four GCs approximately midway between NGC~3379 and NGC~3384, most likely belong to NGC~3379, however they cannot be decisively assigned to either galaxy and are therefore shown as open circles.}
\label{spatplot}
\end{figure}

The spatial distribution of the combined 36 GCs, in relation to
NGC~3379 and NGC~3384, is shown in Fig. \ref{spatplot}. From this
figure, it can be seen that four of the GCs (from our GMOS sample) lie
roughly half way between NGC~3379 and NGC~3384. As NGC~3379 is more
luminous than NGC~3384 (M$_V$ --21.06 vs. --20.64), we might expect
NGC~3379 to have the more populous GC system and indeed this seems to
be the case (Whitlock, Forbes \& Beasley 2003). The average velocity
of the four GCs is $924\pm46$~km/s. The systemic velocity of NGC~3379
is 911~km/s.  For NGC~3384, the systemic velocity is 704~km/s;
however, NGC~3384 has a substantial rotation (it is an S0 galaxy).
The rotation amplitude is about 200~km/s (Fisher 1997) and the side
closest to NGC~3379 is redshifted.  Any rotation in NGC~3379 GC system
is below 2 $\sigma$ significance at all radii and the sample is
consistent with no rotation. In the unlikely situation of the NGC~3384
rotating disk extending to several effective radii, and that the four
GCs in question are part of the NGC~3384 disk population, the expected
mean velocity of NGC~3384 members would be essentially the same as
that for NGC~3379. Therefore, the membership of these four GCs is not
entirely clear. We run the dynamical models below, both including and
excluding these 4 GCs, but our conclusions are not greatly affected
either way.

Ideally, the best estimate of the mass-to-light ($M/L$) variation
would come from models without assumptions about the distribution
function.  The state-of-the-art is to use orbit-based models (e.g.,
Gebhardt \etal 2003; Thomas \etal 2005). However, with only 36 clusters
it is not practical to run these flexible models. For an estimate of
the $M/L$ variation at large radii, we instead employ isotropic models.
While there are theoretical concerns for assuming isotropy at large
radii, these models provide a comparative base to other studies.

For an isotropic model, we only require the second moment of the
projected velocity distribution. The data on the inner parts of the
galaxy come from Gebhardt \etal (2000) and Statler \& Smecker-Hane
(1999).  For the velocity dispersion profile from the GC system we use
a lowess estimator (explained in detail in Gebhardt \etal 1995).  The
lowess technique effectively runs a radial window function through the
data to estimate the velocity squared. It takes the velocity
uncertainties into account directly. We have checked this estimator
against a variety of dispersion estimates, including maximum
likelihood, and find no significant differences.

\begin{figure}
\centerline{\psfig{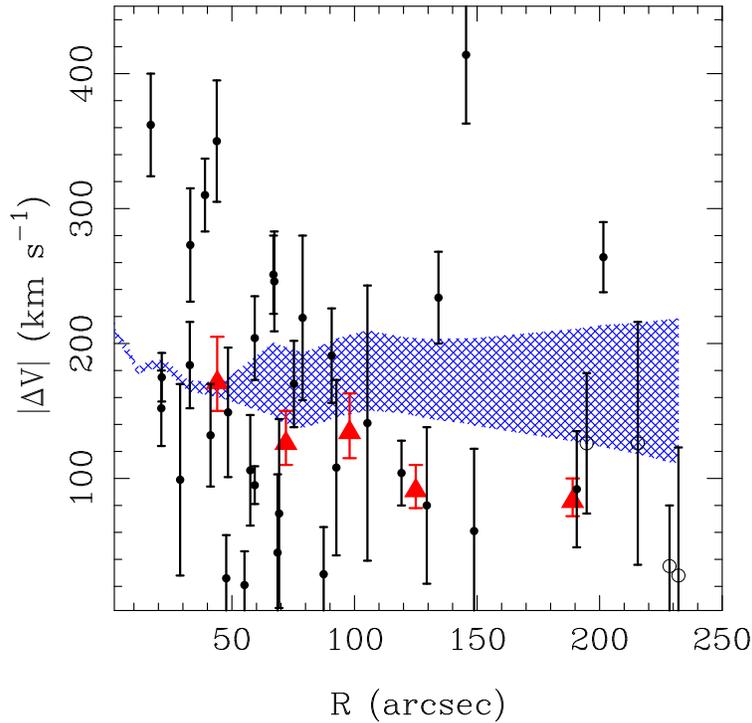}}
\caption{Absolute velocity differences between GCs and the systemic velocity of NGC~3379 (911 km/s) with projected galactocentric radius. The circles are GMOS and VLT data presented here, open circles indicate the four GCs which are possibly associated with NGC~3384.  The red triangles are the binned PN data of Romanowsky \etal (2003).  The hatched blue region shows the 1 $\sigma$ confidence interval for the velocity dispersion from the lowess estimator, including all GCs.}
\label{velrad}
\end{figure}

Fig. \ref{velrad} shows the line-of-sight velocity with projected
radius for the individual GCs.  This allows a comparison of our
unsmoothed data with the lowess estimator velocity dispersion and the
binned Romanowsky \etal (2003) PN data.  The individual data point at
R$\sim$150\arcsec with $\mid\Delta$V$\mid\sim$400 km/s has a relatively
large velocity difference which increases the velocity dispersion at
large radii. If we remove this one point, then the dispersion is
reduced from 166 km/s to 125 km/s at the largest radii (which are the
most uncertain). When compared to the PN dispersion, this results in a
3.5-$\sigma$ difference as opposed to a 4-$\sigma$ difference when
the full sample is used. The technique we are using for the dispersion
has some robustness to outliers built in already, which is why the
change in statistical significance is not particularly strong.

\begin{figure}
\centerline{\psfig{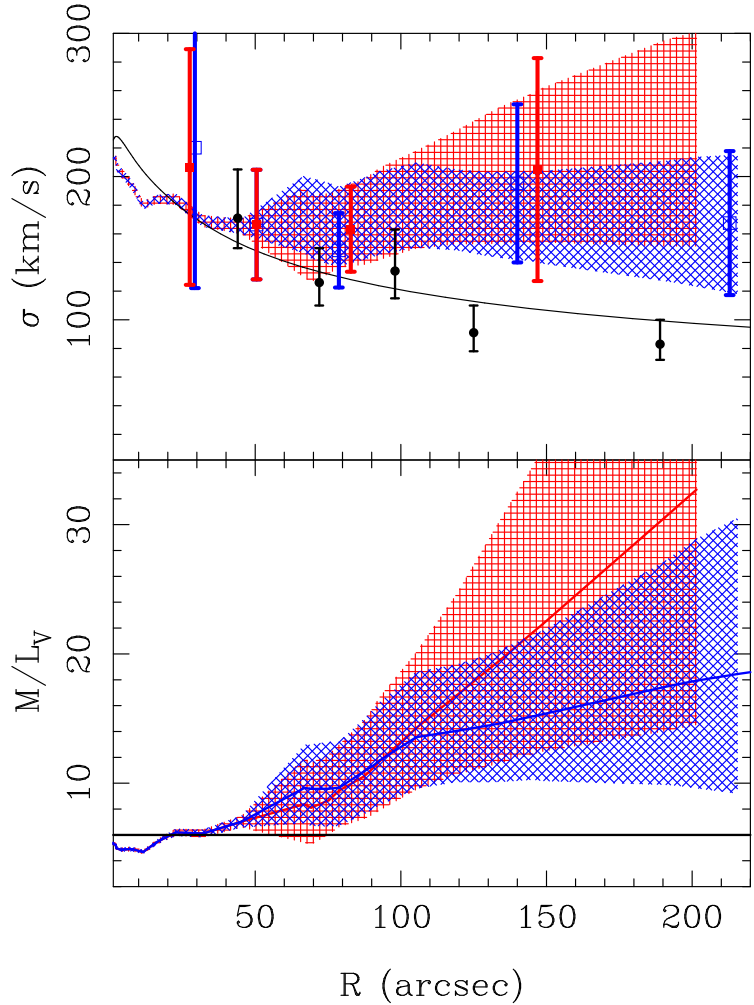}}
\caption{Plots of radial velocity dispersion and M/L ratio using GC
and stellar data (from Gebhardt \etal 2000 and Statler \&
Smecker-Hane 1999). The upper panel shows the velocity dispersion of
the GCs.  Blue open squares and hashing include the
four GCs that that may be associated with NGC~3384, and the red hashing
and filled squares when they are excluded.  The squares show the binned data
and the hashed areas show the range of values using the lowess
estimator (with 1 $\sigma$ errors).  The PN data points from
Romanowsky \etal (2003) (circles) and the expected isotropic profile
with constant M/L (thick line) are also shown. In the lower panel, the
blue line (1 $\sigma$ hatching) shows that when the four ambiguous GCs
are included the M/L ratio rises slowly. The red line (with 1 $\sigma$
hatching) shows that the M/L ratio rises strongly in the outer regions
when the four ambiguous GCs are excluded.  The horizontal line in the
bottom panel is M/L = 6. See text for details. \it{A colour version of
this figure is available in the online journal}.}
\label{DMplots}
\end{figure}

The top panel in Fig. \ref{DMplots} plots the dispersion profile
combining both the stellar and globular cluster data. In the overlap
region the agreement in the dispersion estimates is very good. The
68\% confidence bands come from Monte Carlo simulations as detailed in
Gebhardt \etal (1995).

We use non-parametric, isotropic models as outlined in Gebhardt \&
Fisher (1995). Given the surface brightness profile and velocity
dispersion profile, the spherical Jeans equation can uniquely
determine the mass density profile (and hence the $M/L$ profile)
assuming isotropy. Through the Abel deprojection, the surface
brightness profile uniquely determines the luminosity
density. Similarly, the surface brightness times the projected
velocity dispersion determines the luminosity density times the
internal (3-D) velocity dispersion. From the internal velocity
dispersion, we can derive the mass profile and, hence the $M/L$
profile. We employ a degree of smoothing since the Abel deprojection
involves a derivative (Gebhardt \etal 1996).

The surface brightness profile for the kinematic tracer is important
to characterize well for the dynamical analysis. We use the globular
cluster number density profile from Rhode \& Zepf (2004). Their data
extend from 1.2\arcmin\ to approximately 20\arcmin. The surface
brightness profile for the stellar light comes from Gebhardt \etal
(2000) which extends to 2.8\arcmin. Thus, there is significant overlap
for comparison. In the overlap region, the stellar light is slightly
steeper than the GC profile, and after 2.5\arcmin\ the GC profile
flattens significantly. For the dynamical analysis, one should use the
GC profile only; however, in order to de-project properly, one
requires the central profile which does not exist for the GCs. Thus,
we have to rely somewhat on the stellar profile for the extrapolation
inward. For the dynamical analysis, we use the GC profile from Rhode
\& Zepf into 1.2\arcmin, and then use the stellar profile interior to
that. We have also tried a variety of profiles: using only the cluster
profile with various central extrapolations, using only the stellar
light profile, and using various extrapolations to larger radii. We
find that there is essentially no effect from the large radii
extrapolation. For the small radii extrapolation, the effect on the
projected dispersion is dramatic at small radius but relatively
insignificant at the radii where we have GC velocities. Thus, all of
our tests suggest there is little effect on our overall results from
the surface brightness profile.

The velocity dispersion and M/L ratio are plotted with galactocentric
radius in Fig. \ref{DMplots}. GC data, both including and excluding
the four GCs potentially associated with NGC~3384, are shown. The
heavy solid line in the upper panel in Fig. \ref{DMplots} is the
expected isotropic velocity dispersion profile for a constant M/L
ratio, which is consistent with the Romanowsky \etal (2003) PN
data. In the bottom panel we plot the projected $M/L$ profile,
calculated by the lowess method and using the surface brightness
profile characterization described above. Both cases (including and
excluding the four GCs) shows a rise in the $M/L$ at large radii.
This rise in $M/L$ is stronger when the four GCs are excluded.

Finally, we remind the reader that the above results are based on only
36 GCs as kinematic tracers; clearly such an analysis should be
repeated when larger samples become available.

\section{Discussion}\label{sec_disc} 
 
The reliability of GC parameters derived from Lick indices and our
$\chi^2$ fitting method are affected by several factors, most
importantly the stability of the fit when individual indices are
excluded.  For some GCs it is not possible to find a stable fit that
includes a large number of indices ($>$10).  This is often the case
for the low S/N GCs.  The large index errors of low S/N GCs mean that,
within 3 $\sigma$, the indices can be consistent with a wide range of
ages and metallicities. In this situation the fit can be driven by a
single index which is outside the range spanned by the models, leading
to a fit at the extreme of the SSP models. When this occurs the Monte
Carlo determined errors are large, reflecting the lower accuracy of
these parameters.

To examine the effects of these factors we tested the reliability of the
$\chi^2$ fit metallicities by comparing them to Brodie-Huchra
metallicities.  Fig. \ref{bhplot} shows the good agreement between
the two metallicity measures, with some divergence below [Fe/H]=--2
dex.  We then used {\it g--i} colour as a test of the derived age and
metallicity.  Model predictions for {\it g--i} from BC03, for our $\chi^2$
fit ages and metallicities (derived using TMK04 SSPs), are plotted
against the observed {\it g--i} colours in Fig. \ref{colplots}.  We see
reasonable agreement between the predicted and observed colours, with
the outliers being those low S/N GCs that are at the limits of the models.

Having established the reliability of our $\chi^2$ minimisation fits
for age and metallicity, we find, all of the GCs are consistent with
an old age ($\ga$10 Gyrs) within errors and a spread of metallicities
is seen from [Fe/H] $<$--2 to solar (see Fig.  \ref{agefeplots}).  We
see no discernible age structure.

We next consider the $\alpha$-element abundance ratio parameter
[E/Fe].  Due to the necessary exclusion of the Mg and CN indices (see
Section \ref{sec_age}), the only strongly $\alpha$-element sensitive
indices included in the fits are Mgb and C4668. However, all the plots
of Mgb and C4668 against Fe sensitive indices (Fig. \ref{c46plots})
display the same overall trend, of decreasing $\alpha$-element ratio
with increasing metallicity.  This is highlighted by plotting the
residuals of the model [E/Fe]=+0.3 line to these index pairings.  This
is also seen for [E/Fe] derived from the $\chi^2$ fitting process
(Fig. \ref{fealphaplots}). A similar trend is apparent for GCs in
NGC~1052 (See Fig. 9 of Pierce \etal 2005) and is also noted in Puzia
(\etal 2005). This contrasts with the findings of Carney (1996) for
Galactic GCs in which [$\alpha$/Fe] is largely constant for GCs of all
metallicities. However, the Puzia \etal (2005), NGC~3379 and NGC~1052
results are derived from TMK04 models. These models were constructed
using the results of Trippico \& Bell (1995) which included the
effects of $\alpha$-element variation on stars of solar-metallicity.
The application of the Trippico \& Bell (1995) models to GCs of
[Fe/H]$\sim$--2 may therefore play a part in generating the observed
trend of high $\alpha$-element ratios for GCs at low metallicity.  At
such low metallicity we can see from Fig. \ref{c46plots} that the
indices cannot offer much leverage to differentiate the [E/Fe] values.

However, we note that the $\alpha$-element ratio trend extends up to
the solar metallicity at which the Trippico \& Bell (1995) models were
calculated. Of particular interest are the two GCs with solar
metallicity, [E/Fe]$\sim$--0.25 and ages greater than 10 Gyrs.  While
there is no obvious link, it is interesting to note that X-ray
observations of hot gas in the centre of the NGC 5044 galaxy group
reveal a similar abundance pattern, of a sub-solar $\alpha$-element
ratio at solar metallicity (Buote \etal 2003). A significant number of
type Ia, rather than type II, supernovae are therefore necessary to produce
enough Iron to explain this abundance pattern.  Such a process must
have occurred rapidly due to the old ages of the GCs.

From the measured recession velocities of our sample of 22 GCs, and 14
GCs from Puzia \etal (2004), we find evidence for a significant dark
matter halo. This result does not depend on the inclusion of 4 GCs
which may belong to NGC~3384.  The PN data of Romanowsky \etal (2003)
showed a decrease in velocity dispersion at large radii, suggesting a
minimal DM halo. However, Dekel \etal (2005) recently showed that
stellar orbits in the outer regions of merger-remnant elliptical
galaxies are elongated and that declining PN velocity dispersions do
not necessarily imply a dearth of dark matter.  This demonstrates that
the orbital properties of the kinematic tracers need to be well
understood.  For PN radial orbits need to be included, while for GCs
isotropic models should be sufficient (see Cote \etal 2003), but
caution is still required. Ignoring the differences between dynamical
models, we still find the results from the two studies contradict each
other. 

A two-sided $\chi^2$ test over all radii using all GCs, gives a 2
$\sigma$ difference between the mass profile derived from our GCs
compared to the PN-derived mass model of Romanowsky \etal (2003). This
would indicate borderline statistical difference between our mass
model and the no-dark matter model of Romanowsky \etal. 

To test the significance of the differences between the PN and GC data
we have carried out a Monte Carlo simulation. At the radius of a given
PN binned data point, we sample from a Gaussian distribution using a
mean given by the GC Lowess dispersion estimate and the standard
deviation from the 68\% confidence band at that point. We draw another
Gaussian from the PN value and its uncertainty. We then ask whether
the new PN dispersion is smaller than the new GC dispersion at that
point. For a single measurement, this is only a $\chi^2$ test;
however, generalizing for a range of radii, we determine whether each
of the PN dispersions are smaller than each of the GC dispersions.
This test is rather strict in the sense that if any realisation shows
one PN value above the GC value at that radius then the difference is
not considered. At small radii, the PN and GC dispersions are
consistent, whereas at large radii, the PN dispersions are lower.

We find that the PN dispersions are smaller than the GC dispersions at
$>$97\% confidence beyond an effective radius. Since the PN dispersions
are consistent with mass following light, the GC dispersions suggest a
need for a dark halo. Either the orbital properties of the two tracers
are different (Dekel \etal 2005), or one or both are hampered by low
number statistics.

Two alternate observations suggest the presence of a dark
matter halo around NGC~3379.  {\it Chandra} archival data for NGC 3379
indicate that a hot gaseous halo is present (Fukazawa \etal 2006). The
kinematics of the HI gas ring around the Leo Triplet of NGC~3379,
NGC~3384 and NGC~3389 suggests a mass-to-light ratio of 27 which is
consistent with a DM halo (Schneider 1989).

One possible explanation for the difference between the PN data and
the indication of a normal DM halo, could be that NGC~3379 is a
face-on S0 galaxy, as suggested by Capaccioli \etal (1991).  If a
significant fraction of the PN belong to a disk, this would suppress
the line-of-sight velocity dispersion of the PN relative to that of
the GCs which lie in a more spherical halo.

\section{Conclusions}\label{sec_conc} 
 
We have obtained Gemini/GMOS spectra for 22 GCs around the elliptical
galaxy NGC~3379.  We present ages, metallicities and
$\alpha$-abundance ratios that were derived by applying the
multi-index $\chi^2$ minimisation method of Proctor \& Sansom (2002)
to the SSP models of Thomas \etal (2004).  Metallicity estimates,
derived according to the method of Brodie \& Huchra (1990), agree
closely with those from our $\chi^2$ minimisation method.  We also
find good agreement between the observed colours and those predicted
from our $\chi^2$ minimisation method ages and metallicities. All the
GCs are found to be consistent with old ages, i.e. $\ga$10 Gyr, with
a wide range of metallicities.  This is consistent with the resolved
stellar population work of Gregg \etal (2004) who found the galaxy
stars in the outer regions to be old. We find no evidence for a young
GC sub-population.

The $\alpha$-abundance ratios appear to decrease with increasing
metallicity, however interpretation of this trend is complex and
requires further work.

Using the recession velocities of our 22 GCs and 14 GCs from Puzia
\etal (2004), we measure the projected velocity dispersion of the GC
system and find that it is consistent with being constant with radius
in the outer regions. With this velocity dispersion profile, NGC~3379
appears to posses a dark halo, although we cannot rigorously determine
the dark halo mass. This is in contrast to the earlier claims of
Romanowsky \etal (2003) of a minimal DM halo, based on planetary
nebulae kinematics.

A two-sided $\chi^2$ test over all radii using all GCs, gives a
2$\sigma$ difference between the mass profile derived from our GCs
compared to the PN-derived mass model of Romanowsky \etal (2003). This
would indicate borderline statistical difference between our mass
model and the no-dark matter model of Romanowsky \etal However, if we
restrict our analysis to radii beyond one effective radius and test if
the GC velocity dispersion is consistently higher, then we determine a
$>3\sigma$ difference between the mass models, and hence favor the
conclusion that NGC 3379 does indeed have dark matter at large radii
in its halo.

\section{Notes added in Proof}

We emphasise that the mass model we compare to, and find discrepancy
with, is an isotropic constant M/L model, which is the simplest
interpretation of the PN data implying no DM.  This is not the same as
the preferred mass model presented in Romanowsky \etal (2003), which
was an orbit model including some dark matter.  For detailed
comparison to such a model, the dynamical characteristics and
projection effects for both the GCs and PN would need to be taken into
account, which is beyond the scope of this work.

Our attention was drawn to the paper by Samurovic \& Danziger
(2005). Their analysis of the X-ray halo of NGC~3379 leads to
predictions of both M/L and velocity dispersion with radius that are
consistent with those from our GC analysis For example, see their
Figs. 10 and 13 and their statement "We note that beyond
120$^{\prime\prime}$ (2.2 R$_e$) a discrepancy between PNe estimates and
X-ray estimate occurs".

\section{Acknowledgments}\label{sec_ack} 
 
We thank the Gemini support staff for help preparing the slit mask. We
thank S. Brough, A. Romanowsky and S. Samurovic for useful comments.
DF thanks the ARC for its financial support.  SEZ acknowledges support
for this work in part from the NSF grant AST-0406891 and from the
Michigan State University Foundation. This research was supported in
part by a Discovery Grant awarded to DAH by the Natural Sciences and
Engineering Research Council of Canada (NSERC).

These data were based on observations obtained at the
Gemini Observatory, which is operated by the Association of
Universities for Research in Astronomy, Inc., under a cooperative
agreement with the NSF on behalf of the Gemini partnership: the
National Science Foundation (United States), the Particle Physics and
Astronomy Research Council (United Kingdom), the National Research
Council (Canada), CONICYT (Chile), the Australian Research Council
(Australia), CNPq (Brazil), and CONICET-Agencia Nac. de Promocion
Cientifica y Tecnologica (Argentina).  The Gemini program ID is
GN-2003A-Q22. This research has made use of the NASA/IPAC
Extragalactic Database (NED), which is operated by the Jet Propulsion
Laboratory, Caltech, under contract with the National Aeronautics and
Space Administration.

\end{document}